\documentclass[a4paper,12pt]{article}

\usepackage{graphicx,amssymb,bm,latexsym,color,mathrsfs,textcomp}
\usepackage{amsmath}
\makeatletter

\@addtoreset{equation}{section}
\makeatother

\begin{document}

\baselineskip=17.5pt
\begin{titlepage}

\begin{center}
\vspace*{10mm}

{\large\bf Lorentz breaking terms from Einstein gravity
}%
\vspace*{12mm}

Takayuki Hirayama\footnote{hirayama@isc.chubu.ac.jp}
\vspace{4mm}

{\it College of Engineering, Chubu University,
\\
1200 Matsumoto, Kasugai, Aichi, 487-8501, Japan}\\%
\vspace*{12mm}

\begin{abstract}\noindent%
We construct an action which is invariant under the foliation preserving diffeomorphism from the Einstein Hilbert action. Starting from the Einstein Hilbert action, we introduce the gauge invariance under the anisotropic rescaling by using the St\"uckelberg method. We then introduce the gauge field corresponds to the anisotropic rescaling, and the St\"uckelberg field turns out to be the Nambu-Goldstone boson. The Nambu-Goldstone boson, however, is not completely eaten by the massive gauge field and the Nambu-Goldstone boson can be integrated out from the action. Then the resultant action is a Horava-Lifshitz type action which contains $R^3$ term.
\end{abstract}

\end{center}

\end{titlepage}

\newpage

\section{Introduction}

It have been known that the quantum theory of general relativity is non renormalizable~\cite{'tHooft:1974bx} and there are many attempts to construct a renormalizable quantum gravity. Many different approaches seem suggesting that at high energy the dimension of space time approach to two in some sense. One proposal by Horava~\cite{Horava:2009uw} is that if Lorentz symmetry is abandoned and the dispersion relation between the energy $\omega$ and momentum $k$ is given $\omega^2= k^6$ at high energy, the spectral dimension becomes two in four dimensions~\cite{Horava:2009if} suggesting that the Horava-Lifshitz gravity theory becomes renormalizable.

Although the diffeomorphism invariance is broken, the Horava-Lifshitz gravity, instead, requires the invariance under the foliation preserving diffeomorphism at high energy, and the theory is characterized by the invariance under the anisotropic space time rescaling with the dynamical critical exponent $z$,
\begin{align}
 t \rightarrow b^z t, \hspace{3ex} x^i \rightarrow b x^i,
\end{align}
where $b$ is a scaling parameter, and $t$ and $x^i$ are time and spacial directions respectively. The action consists of second order time derivative kinetic terms and various higher order (up to $2z$ order) spacial derivative potential terms. Since the theory does not have the full differmorphism invariance, a scalar mode, in addition to spin two modes, generically appears in the metric~\cite{Charmousis:2009tc}. Therefore there are technical difficulties to successfully recover Einstein gravity at low energy. 

Although the higher order spacial derivative terms are allowed by the symmetry, we can ask how these terms apear. In this paper, we try to discuss the origin of those terms. We will derive some of those terms by integrating out the Nambu-Goldstone mode associated with the breaking of the anisotropic rescaling. To do this, at first, we modify the Einstein Hilbert action such that the action has the foliation preserving difeomorphism and anisotropic rescaling invariance, but has only second order spacial derivatives. After that we show that the Nambu-Goldstone mode can be integrated out in an unitary gauge and the resultant action is a Horava-Lifshitz type action which contains higher order spacial derivative terms.

\section{The St\"uckelberg method}

When an action has a global symmetry, for example $U(1)$ global symmetry, we can gauge the global symmetry by introducing the corresponding gauge fields $A_\mu$ and replacing the derivatives with the covariant derivatives. The transformation law for the gauge fields is, $A_\mu \rightarrow A_\mu -\partial_\mu \lambda$ for $U(1)$ with the transformation parameter $\lambda$. We can also introduce the kinetic term for the gauge field, $-(1/4)F^{\mu\nu}F_{\mu\nu}$.

In the same way, we construct a gravity action which is invariant under the anisotropic rescaling in this section. We start with the four dimensional Einstein gravity,
\begin{flalign}
S_0 &=\!\! \int \!\! d^4x \sqrt{-g} \big[ R^{(4)} -2\Lambda \big],
\end{flalign}
where the signature of the metric is $(-,+,+,+)$. If the cosmological constant is positive (negative), the dS (AdS) space, $R_{\mu\nu}=\Lambda g_{\mu\nu}$, is a solution of Einstein equations. This action is invariant under the diffeomorphism transformation, $x'^\mu=x^\mu-\xi^\mu$, ($\mu=0,1,2,3$),
\begin{flalign}
 \delta_{\xi} g_{\mu\nu} &= (\partial_\mu \xi^\rho) g_{\rho\nu} 
 +(\partial_\nu \xi^\rho) g_{\mu\rho} + \xi^\rho \partial_\rho g_{\mu\nu} 
 = \nabla_\mu\xi_\nu+\nabla_\nu\xi_\mu.
\end{flalign}

\noindent
It takes two steps to obtain the action which is invariant under the anisotropic rescaling with the dynamical critical exponent $z$ and the parameter $b$,
\begin{align}
 t \rightarrow b^z t, \hspace{3ex} x^i \rightarrow b x^i,
\end{align}
where $t=x^0$, and $i=1,2,3$ are the indices for the three spacial directions.

First we introduce the St\"uckelberg~\cite{stuc} field $X$ in the following way,
\begin{align}
 g_{\mu\nu} = \widetilde{g}_{\mu\nu} X^2.
\end{align}
We substitute this redefinition in the action, and we obtain
\begin{align}
 S_0 &= \int  \!\! d^4 x \sqrt{-\widetilde{g}} \big[ X^2 \widetilde{R}^{(4)} + 6 (\widetilde{\partial}X)^2 -2X^4\Lambda \big],
\end{align}
which is nothing but the Einstein action with a conformally coupled scalar field $X$.
We then use the ADM decomposition,
\begin{flalign}
 \widetilde{g}_{00} &= -(\widetilde{N}^2-N_iN^i), \hspace{3ex}
 \widetilde{g}_{0i} = N_i, \hspace{3ex}
 \widetilde{g}_{ij} = \gamma_{ij},
 \\
 \widetilde{g}^{\mu\nu} &= \frac{1}{\widetilde{N}^2}
 \left( \begin{array}{cc}
 -1 &N^i
 \\
 N^j& \widetilde{N}^2\gamma^{ij} -N^iN^j
 \end{array}\right) ,
 \\
ds^2 &= -\widetilde{N}^2 dt^2 + \gamma_{ij} ( dx^i + N^i dt) ( dx^j + N^j dt) ,
\end{flalign} 
where $N^i = \gamma^{ij} N_j$. The action becomes
\begin{flalign}
 S_0 &=\!\! \int \!\! d^4x \sqrt{\gamma} \widetilde{N} \big[
 X^2(\widetilde{K}_{ij}\widetilde{K}^{ij} - \widetilde{K}^2) +X^2R^{(3)} 
 -\frac{6X^2}{\widetilde{N}^2}(\partial_t \ln X - N^i\partial_i\ln X)^2
 \nonumber
 \\
 &\hspace{6ex}
 +6X^2\gamma^{ij}(\partial_i \ln X)(\partial_j \ln X)
 -2X^4\Lambda \big]
  \nonumber
  \\
 &\hspace{6ex}
 -2X^2 \partial_t ( \sqrt{\gamma} \widetilde{K}) +2\sqrt{\gamma}X^2\nabla_i (\widetilde{K}N^j) -2X^2 \sqrt{\gamma}\Delta \widetilde{N}
 ,
 \\
 \widetilde{K}_{ij} &=\frac{1}{2\widetilde{N}}( -\partial_t \gamma_{ij} + \nabla_i N_j + \nabla_j N_i ), \hspace{3ex}
 \widetilde{K} = \widetilde{K}_{ij}\gamma^{ij},
\end{flalign}
where the indices are contracted with $\gamma_{ij}$. $R^{(3)}$ and $\nabla_i$, ($\Delta=\nabla^i\nabla_i$), are the Ricci scalar and the covariant derivatives constructed from $\gamma_{ij}$.
The diffeomorphism transformations are then,
\begin{flalign}
 \delta_\xi \gamma_{ij} &=  \delta^{(3)} \gamma_{ij} 
 +N_i\partial_j \xi^t+N_j\partial_i \xi^t +\xi^t\partial_t \gamma_{ij},
 \\
 \delta_\xi N_i &= \gamma_{ij}\partial_t \xi^j +\delta^{(3)} N_i+ N_i\partial_t\xi^t
 +\xi^t\partial_tN_i-(\widetilde{N}^2-N_jN^j)\partial_i\xi^t,
 \\
 \delta_\xi \widetilde{N} &= \delta^{(3)} \widetilde{N} + \xi^t\partial_t \widetilde{N} + \widetilde{N}\partial_t \xi^t
 -\widetilde{N}N^i\partial_i \xi^t,
 \\
 \delta_\xi X &= \delta^{(3)} X +\xi^t\partial_t X,
 \\
 &~~~~
 \delta^{(3)} \gamma_{ij} = (\partial_i \xi^k) \gamma_{kj} 
 +(\partial_j \xi^k) \gamma_{ik} + \xi^k \partial_k \gamma_{ij} 
 =\nabla_i \xi_j +\nabla_j\xi_i,
 \\
 &~~~~
 \delta^{(3)} N_i = N_j\partial_i\xi^j+\xi^j\partial_jN_i,\hspace{3ex}
 \delta^{(3)} \widetilde{N} = \xi^i\partial_i \widetilde{N}, \hspace{3ex}
 \delta^{(3)} X = \xi^i\partial_i X.
\end{flalign}

The second step is the redefinition of $\widetilde{N}$,
\begin{flalign}
 \widetilde{N} = N X^{z-1}.
\end{flalign}
After this replacement, we have
\begin{flalign}
 S_0 &=\!\! \int \!\! d^4x \sqrt{\gamma} N \big[
 X^{-z+3}(K_{ij}K^{ij} - K^2) +X^{z+1}R^{(3)} 
  -\frac{6X^{-z+3}}{N^2}(\partial_t \ln X - N^i\partial_i\ln X)^2
 \nonumber
 \\
 &\hspace{2ex}
 +6X^{z+1}\gamma^{ij}(\partial_i \ln X)(\partial_j \ln X)
 -2X^{z+3}\Lambda \big]
 \nonumber
  \\
 &\hspace{2ex}
 +4\sqrt{\gamma}X^{-z+3}K ( \partial_t -N^i\partial_i)\ln X
 -4\sqrt{\gamma}N X^{z+1}\big\{2 (\partial^i \ln X)( \partial_i \ln X) + \Delta \ln X \big\}
 , \label{q1}
 \\
 K_{ij} &=\frac{1}{2N}( -\partial_t \gamma_{ij} + \nabla_i N_j + \nabla_j N_i ).
\end{flalign}
Here we used the integration by parts in the last line in~\eqref{q1}.
So far we have only redefined the fields and thus we have not changed the theory at all. And this action, of course, is invariant under the diffeomorphism transformations which are now,
 \begin{flalign}
 \delta_\xi \gamma_{ij} &= \delta^{(3)} \gamma_{ij} 
 +N_i\partial_j \xi^t +N_j\partial_i \xi^t +\xi^t\partial_t \gamma_{ij},
 \\
 \delta_\xi N_i &= \gamma_{ij}\partial_t \xi^j +\delta^{(3)} N_i+ N_i\partial_t\xi^t
 +\xi^t\partial_tN_i-(X^{2z-2}N^2-N_jN^j)\partial_i\xi^t,
 \\
 \delta_\xi N &= \delta^{(3)} N + \xi^t\partial_t N + N\partial_t \xi^t
 -NN^i\partial_i \xi^t,
 \\
 \delta_\xi X &= \delta^{(3)} X +\xi^t\partial_t X.
\end{flalign}
In addition to the diffeomorphism invariance, this action is invariant under the anisotropic rescaling,
\begin{align}
 N \rightarrow NY^z, \hspace{3ex}
 N_i \rightarrow N_i Y^2, \hspace{3ex}
 \gamma_{ij} \rightarrow \gamma_{ij} Y^2,
 \hspace{3ex}
 X \rightarrow Y^{-1}X.
 \label{2}
\end{align}
This anisotropic rescaling is equivalent with the anisotropic space time rescaling in the following way,
\begin{align}
 t \rightarrow b^z t, \hspace{3ex} x^i \rightarrow b x^i, \hspace{3ex}
 N_i \rightarrow b^{-z+1} N_i, \hspace{3ex}
 X\rightarrow b^{-1}X.
 \label{st1}
\end{align}
This is achieved from the combination of the general coordinate transformation, $t'=b^zt$ and $x'^i=bx^i$, and the anisotropic rescaling with $Y=b$.
Since $N_i$ and $X$ transform non trivially, this is not exactly same as the anisotropic rescaling of space and time. We come to this point later.

The infinitesimal transformation of the anisotropic rescaling~\eqref{2}, $Y=1+\lambda$, is
\begin{eqnarray}
 \delta_\lambda N = z\lambda N, \hspace{3ex}
 \delta_\lambda N_i = 2\lambda N_i, \hspace{3ex}
 \delta_\lambda \gamma_{ij} = 2\lambda \gamma_{ij}, \hspace{3ex}
 \delta_\lambda X = -\lambda X .
\end{eqnarray}
Thus, $\partial_\mu \ln X$ transforms as a corresponding gauge field does,
\begin{align}
 \partial_\mu \ln X \rightarrow \partial_\mu \ln X - \partial_\mu \lambda.
\end{align}
Therefore a heuristic way of constructing a gauge theory is introducing the gauge field $A_\mu$ by replacing $\partial_\mu \ln X$ with $A_\mu$. After the replacement, we finally have,
\begin{flalign}
 S_1
 &=\!\! \int \!\! d^4x \sqrt{\gamma} N \big[
 X^{-z+3}{\cal K} - X^{z+1}V_{z+1} - X^{z+3} V_{z+3} \big],
 \label{q2}
 \\
 {\cal K}&= K_{ij}K^{ij} - K^2
 -\frac{6}{N^2}(A_t - N^iA_i)^2
  +4\frac{K}{N} ( A_t - N^iA_i),
 \\
 &~~~~
 K_{ij} =\frac{1}{2N}( -\partial_t \gamma_{ij} + \nabla_i N_j + \nabla_j N_i ),
 \\
 V_{z+1}&= -R^{(3)} 
 +2\gamma^{ij}A_iA_j
 +4\nabla^i A_i,
 \hspace{3ex}
 V_{z+3} = 2\Lambda.
\end{flalign}
The transformations under the differmorphism and anisotropic rescaling for $A_\mu$ are given
\begin{align}
 \delta_\xi A_t &= A_i\partial_t \xi^i +\delta^{(3)} A_t + A_t \partial_t \xi^t +\xi^t \partial_t A_t,
 \\
 \delta_\xi A_i &=  \delta^{(3)} A_i + A_t \partial_i \xi^t +\xi^t \partial_t A_i,
 \\
 &~~~~
 \delta^{(3)} A_t = \xi^i\partial_i A_t, \hspace{3ex}
 \delta^{(3)} A_i = A_j\partial_i \xi^j + \xi^j\partial_j A_i,
 \\
 \delta_\lambda A_\mu &= -\partial_\mu \lambda,
\end{align}
where $\delta_\xi A_t$ and $\delta_\xi A_i$ are easily computed from $\delta_\xi A_\mu= A_\nu \partial_\mu \xi^\nu + \xi^\nu\partial_\nu A_\mu$.
We can think the action~\eqref{q2} is an anisotropic rescaling gauge theory in an unitary gauge where the Nambu-Goldstone boson $X$ is eaten by the massive gauge fields $A_\mu$. However, the Nambu-Goldstone boson is not completely eaten by the gauge fields.
We notice that this replacement changes the theory. For example, some of the symmetries are lost, and dS${}_4$ (AdS${}_4$) space for the positive (negative) cosmological constant is no longer a solution. Instead, as we will see in the next section that Minkowski space is a solution independent of the value of the cosmological constant.
We can check explicitly which of the symmetries remain, and find that the anisotropic rescaling symmetry remains, but the diffeomrophism invariance is broken down to the foliation preserving diffeomorphism invariance. The foliation preserving diffeomorphism is generated by $\xi^i(t,x)$ and $\xi^t(t)$. Under the transformation of the anisotropic rescaling, the $X^{-z+3}$, $X^{z+3}$ and $X^{z+1}$ parts are independently invariant. The transformation laws for the three dimensional part of the diffeomorphism generated by $\xi^i(t,x)$ are same as those of three dimensional diffeomorphism, $\xi^i(t,x)=\xi^i(x)$, except that $\delta_\xi N_i$ has the additional term $\gamma_{ij}\partial_t \xi^i$ and $\delta_\xi A_t$ has the additional term $A_i\partial_t \xi^i$. Because of these, the combination $\partial_t \gamma_{ij} -\nabla_i N_j - \nabla_j N_i$ and $A_t-N^iA_i$ are invariant, and thus the action is invariant. Under the diffeomorphism transformation generated by $\xi^t(t,x)$, one can check explicitly that the action is not invariant and the deviation of the action $\delta S$ is proportional to $A_\mu - \partial_\mu \ln X$. However since the action does not explicitly depend on time, it is easily checked the action is still invariant under the diffeomorphism transformation by $\xi^t(t,x)=\xi^t(t)$. Therefore the action~\eqref{q2} has the foliation preserving diffeomorphism and anisotropic rescaling invariance.

We give two comments. We can add the kinetic term for the gauge field, $\sqrt{-\widetilde{g}} \widetilde{g}^{\mu\rho}\widetilde{g}^{\nu\tau}F_{\mu\nu}F_{\rho\tau}$, and this term introduces additional $X^{z-1}$ and $X^{-z+1}$ terms. 
We can add 
$\sqrt{-\widetilde{g}} X^{2}\widetilde{g}^{\mu\nu}Y_\mu(A_\nu-\partial_\nu \ln X)$ in the action where $Y_\mu$ fields are Lagrange multipliers in order to keep the resultant action equivalent with the Einstein Hilbert action.

\section{Higher order spacial derivative terms}

In the resultant action~\eqref{q2}, $X$ field appears only in front of ${\cal K}$, $V_{z+1}$ and $V_{z+3}$ and thus can be integrated out. We do this for the case $z=3$ in which we are most interested.

For $z=3$, the action becomes simpler,
\begin{flalign}
 S_1 &=\!\! \int \!\! d^4x \sqrt{\gamma} N \big[
 {\cal K} - X^{4}V_{4} - X^{6} V_{6} \big],
 \label{f1}
 \\
 {\cal K}&= K_{ij}K^{ij} - K^2
 -\frac{6}{N^2}(A_t - N^iA_i)^2
  +4\frac{K}{N} ( A_t - N^iA_i),\\
 V_4&= -R^{(3)} 
 +2\gamma^{ij}A_iA_j
 +4\nabla^i A_i,
 \hspace{3ex}
 V_6 = 2\Lambda.
\end{flalign}
The equation of motion for $X$ is
\begin{align}
 4X^3 V_4 +6X^5 V_6 =0.
\end{align}
This has a non trivial solution,
\begin{align}
 X &= \sqrt{ -\frac{2V_4}{3V_6}}.
\end{align}
We notice that the value of action is real, even when $X$ is pure imaginary. Therefore this non trivial solution is allowed no matter what the sign of $V_4$ and that of $V_6$ are.
We substitute this non trivial solution into the action, and we obtain
\begin{flalign}
 S_1 &=\!\! \int \!\! d^4x \sqrt{\gamma} N \big[
 {\cal K} - \frac{4(V_4)^3}{27(V_6)^2}\big].
 \label{f2}
\end{flalign}
We can easily see that $(R^{(3)})^3$ term from $(V_4)^3$ has higher order spacial derivative terms. However these are interaction terms and are not the higher order spacial derivative bi-linear terms for the spin two graviton around Minkowski space. One can check from the action~\eqref{f1} or~\eqref{f2} that Minkowski space, i.e. $\gamma_{ij}=\delta_{ij}$, $N=$ constant and $A_\mu=0$, is always a solution.

Here we quote the Horava-Lifshitz gravity action with $z=3$ in~\cite{Barvinsky:2015kil},
\begin{eqnarray}
S_{Horava} \!\!\! &=& \!\!\! \int \!\! d^4x \sqrt{\gamma}N\big[K_{ij}K^{ij}-\lambda K^2 -V\big],
\\
K_{ij} \!\!\! &=& \!\!\! \frac{1}{2N}( - \partial_t \gamma_{ij} +\nabla_i N_j + \nabla_j N_i),
\\
V \!\!\! &=& \!\!\! 2\Lambda -\eta R + \mu_1 R^2 + \mu_2 R_{ij}R^{ij} 
\nonumber
\\&&
+\nu_1R^3 + \nu_2 R R_{ij}R^{ij} +\nu_3 R^i_jR^j_kR^k_i +\nu_4 \nabla_iR\nabla^iR +\nu_5 \nabla_iR_{jk}\nabla^iR^{jk},
\end{eqnarray}
where $R$ and $R_{ij}$ are the Ricci scalar and Ricci tensor constructed from $\gamma_{ij}$. The dispersion relations of spin two and zero modes are,
\begin{eqnarray}
 \omega_2^2 \!\!\! &=& \!\!\! \eta k^2 +\mu_2 k^4 +\nu_5 k^6,
 \\
 \omega_0^2 \!\!\! &=& \!\!\! \frac{1-\lambda}{1-3\lambda} \big( -\eta k^2 + ( 8\mu_1+3\mu_2)k^4 + ( 8\nu_4+3\nu_5) k^6 \big),
\end{eqnarray}
where $\omega_2$ and $\omega_0$ are the energy of spin two and zero modes and $k$ is the momentum.
Thus, what we have successfully obtained are limited terms of Horava gravity theory.

We can also integrate out $A_t$ which appears in ${\cal K}$. The equation of motion for $A_t$ tells
\begin{align}
 A_t-N^iA_i = \frac{NK}{3},
\end{align}
and ${\cal K}$ becomes
\begin{align}
 {\cal K} = K_{ij}K^{ij}-\frac{1}{3}K^2.
\end{align}
Since the coefficient in front ot $K^2$ is $-1/3$, the kinetic term for spin zero part of the graviton does not appear.
In fact, we can solve the equations of motion for $A_i$ from the action~\eqref{f1} and $A_i=\partial_i \ln N+ 4\partial_i \ln X$. We substitute this solution and have spacial derivative terms for $X$. However we still do not have kinetic term for $X$ and then can integrate out $X$ field. It will be interesting to study the resultant action.

\section{Discussion}

We have derived some of higher order spacial derivative terms as a result of spontaneous breaking of anisotropic rescaling symmetry. The breaking is occurred once we take a background, for example Minkowski space.
We do not obtain $\nabla_iR_{jk}\nabla^iR^{jk}$ term which give $k^6$ contribution to the dispersion relation of the spin two graviton. This is because $\nabla_iR_{jk}\nabla^iR^{jk}$ can not be rewritten a polynomial of scalar field $X$.
One may think what terms will be introduced if we introduce matter fields in the action. Once we introduce a scalar field $\phi$, there are the additional terms $\gamma^{ij}(\partial_i\phi)(\partial_j \phi)$ to $V_4$, and $m^2\phi^2$ to $V_6$, respectively. Then we have $(\partial_i\phi)^6$ interacting term from $(V_4)^3/(V_6)^2$.
Since the theory~\eqref{q2} is no longer same as the original theory, the dS (AdS) space is no longer a solution. We have to resolve the equations of motion to study the background and perturbation theory of the action~\eqref{f2}. Although Minkowski space is a solution, the solution space is more limited compared with the one in the original theory, since the gauge fields $A_\mu$ also do not have kinetic terms and are not dynamical. Therefore we come to think about the kinetic term for $A_\mu$. The kinetic term which does not break the symmetries of the theory is,
\begin{align}
 &\int \!\!  d^4x \sqrt{-\widetilde{g}} \Big[-\frac{1}{4}\widetilde{g}^{\mu\nu}\widetilde{g}^{\rho\sigma} F_{\mu\rho}F_{\nu\sigma} \Big]
 \nonumber
 \\
 &~~~= \int \!\! d^4x \sqrt{\gamma}N \Big[
 X^{-z+1}\big\{
 \frac{1}{2N^2} \gamma^{ij}F_{ti}F_{tj} - \frac{1}{N^2} N^j \gamma^{ik} F_{ti}F_{jk} \big\}
 +X^{z-1} \big\{ -\frac{1}{4} \gamma^{ij}\gamma^{kl} F_{ik}F_{jl} \big\}
 \Big],
\end{align}
where $F_{\mu\nu}=\partial_\nu A_\nu - \partial_\nu A_\mu$ as usual. Then the kinetic term will induce more complex potential terms and it is interesting to study the solutions of equations of motion.

We can keep the theory~\eqref{q2} equivalent with Einstein gravity by introducing the following term, 
\begin{align}
 &\hspace{-6ex}
 \int \!\!  d^4x \sqrt{-\widetilde{g}} X^{2}\widetilde{g}^{\mu\nu}Y_\mu(A_\nu-\partial_\nu \ln X)
 \nonumber\\
 &=\int \!\!  d^4x \sqrt{\gamma}N \Big[
 -\frac{X^{-z+3}}{N^2}
 (Y_t-N^iY_i) ( A_t - N^j A_j  - \partial_t \ln X +N^j\partial_j \ln X)
 \nonumber\\
 &~~~~~~~~~~~~~~~~~~~~
 + X^{z+1}\gamma^{ij}Y_i(A_j-\partial_j\ln X) \Big],
\end{align}
where $Y_\mu$ are Lagrange multipliers, and the action is invariant under the full diffeomorphism by giving appropriate transformations for $Y_\mu$. It is also interesting to study the effects of this term. 

We could not induce the terms which contribute to $k^6$ in the dispersion relation. 
If we are allowed to introduce spin two graviton mass terms~\cite{Fierz:1939ix}  (for review~\cite{derham}) in the action and fine tune the cosmological constant to be zero, $(V_4)^3/(V_6)^2$ term is schematically written as,
\begin{align}
 (V_4)^3/(V_6)^2\sim (\partial_i h^{TT}_{jk})^6/(m^2(h^{TT}_{jk})^2)^2\sim k^6(h^{TT}_{jk})^2,
\end{align}
where $h^{TT}_{jk}$ denotes the spin two graviton mode. Therefore we have higher order spacial derivative bi-linear terms.

Under the space time rescaling in~\eqref{st1}, $N_i$ and $X$ transform. If we further redefine $N_i= N'_i X^{z-1}$, $N'_i$ does not transform under the space time anisotropic rescaling. Then
\begin{align}
 K_{ij} &= 
 \frac{1}{2N}\Big[-\partial_t \gamma_{ij} + X^{z-1}\big\{\nabla_i N'_j + \nabla_j N'_i +(z-1)( (\partial_i \ln X)N'_j + (\partial_j \ln X) N'_i) 
 \big\}\Big],
 \\
 &\rightarrow \frac{1}{2N}\Big[-\partial_t \gamma_{ij} + X^{z-1}\big\{\nabla_i N'_j + \nabla_j N'_i +(z-1)(A_iN'_j + A_j N'_i \big)\big\}\Big],
\end{align}
and all the $N_i$ are replaced with $N'_iX^{z-1}$ in the action~\eqref{q2}. Therefore only $X$ field remains transforming. 
This $X$ field naturally appears in the theory which admits the Lifshitz geometry~\cite{Koroteev:2007yp}. The metric of the Lifshitz geometry is,
\begin{align}
 ds^2 &= - \frac{dt^2}{r^{2z}} + \frac{\gamma_{ij}dx^idx^j}{r^2} + \frac{dr^2}{r^2}.
 \label{lif}
\end{align}
This metric is invariant under the anisotropic space time rescaling, $t\rightarrow b^zt$ and $x^i\rightarrow bx^i$ provided with $r\rightarrow br$.
We then replace $r$ in the metric~\eqref{lif} by the radion field $R$ and $R$ acts as $X^{-1}$. This is clear if we compactify $r$ direction and see the form of the effective action~\cite{hira}. The radion field has an ordinary kinetic term, and has a stable non trivial minimum when the sign of $V_4$ is negative and that of $V_6$ is positive for $z=3$. This is realized if $\gamma_{ij}$ in~\eqref{lif} describes dS${}_4$ space for instance. In this case, when we consider the perturbation of graviton or matter fields around the background, we have higher order spacial derivatives as higher order perturbations from $R^{-4}V_4$ and $R^{-6}V_6$ terms once we take into account of the back reaction to $R$ field. In other words, once we firstly integrate out $R$ field in the functional integral, the higher spacial derivative terms are induced.

We expect there are many directions to be explored by using the method in this paper.

\end{document}